\documentclass[11pt,a4paper]{article}
\usepackage{color}
\usepackage{amsfonts}
\usepackage{amssymb}
\usepackage{graphicx}
\usepackage{amsmath}
\usepackage{enumerate}
\usepackage{color}

\RequirePackage{mathrsfs} \RequirePackage[sc]{mathpazo}
\RequirePackage{wasysym} \RequirePackage{setspace} \textheight=650pt
\newcommand{\be}{\begin{equation}}
\newcommand{\ee}{\end{equation}}
\newcommand{\bea}{\begin{eqnarray}}
\newcommand{\eea}{\end{eqnarray}}
\begin{document}

\title{  { \begin{flushright}
{\normalsize\small } \end{flushright} }\bf
    Fractional Quantum Hall Filling Factors from String Theory using   Toric Geometry  }
\author{  A. Belhaj$^1$, Z. Benslimane$^{2}$,   M. El  Idrissi$^1$, B. Manaut$^1$,
 A. Sebbar$^3$, M.B. Sedra$^{2}$ \hspace*{-15pt} \\
 {\small $^{1}$D\'epartement de Physique, LIRST,  Facult\'e
Polydisciplinaire,
Universit\'e Sultan Moulay Slimane}\\
{\small B\'eni Mellal, Morocco}\\{\small $^{2}$ D\'{e}partement de
Physique, LHESIR, Facult\'{e} des Sciences, Universit\'{e} Ibn
Tofail,
 K\'{e}nitra, Morocco} \\
{\small $^{3}$Department of Mathematics and Statistics, University
of Ottawa}\\{\small 585 King Edward Ave.,
 Ottawa, ON, Canada, K1N 6N5} }   \date{}\maketitle
\begin{abstract}
Using toric   Cartan matrices as abelian  gauge charges, we present
a class of stringy fractional quantum  Hall effect (FQHE) producing
some recent experimental observed filling factor values. More
precisely, we derive the corresponding Chern-Simons type models from
M-theory compactified on four complex dimensional hyper-K\"{a}hler
manifolds $X^4$. These manifolds, which are viewed as target spaces
of a particular  $N=4$ sigma model in two dimensions, are identified
with the  cotangent bundles over intersecting 2-dimensional toric
varieties $V_i^2$ according to   toric Cartan matrices. Exploring
results of string dualities, the presented  FQHE can be obtained
from  D6-banes wrapping  on  such  intersecting  toric varieties and 
interacting  with   R-R gauge fields. This string theory realization
provides  a geometric interpretation of the filling factors in terms
of toric and Euler characteristic topological data
of the compactified geometry. Concretely,  explicit bilayer models  are worked out in some details. 
\end{abstract}

\newpage
\tableofcontents \thispagestyle{empty} \newpage \setcounter{page}{1}
\newpage
\section{Introduction}
Considerable   efforts have been devoted to study connections
between the quantum theory of condensed matter physics and higher
dimensional supergravity models embedded in  ten dimensional  string
theory \cite{1,2,200,3}.  Concretely, the fractional quantum Hall
effect (FQHE) describing the behavior of  the  two dimensional
electronic system in  the presence of an external magnetic field has
been subject to some interest not only because of its experimental
results related to material physics \cite{4},   nor  its connection
with the recent developments in type II superstrings  and M-theory
compactifications\cite{5,6,7,8,9} but also because of   its relation
to topological calculation using knot theory \cite{90}.\newline

It is  recalled  that  the first proposed  class  of the fractional
quantum states was  proposed  by
Laughlin and  it is  characterized by the filling factor $\nu _{L}=\frac{1}{%
k}$ where $k$ is an even integer for a bosonic electron and an odd
integer for a fermionic electron \cite{10,11}. At low energy, the
corresponding  physical system   can be described by a 3-dimensional
 Chern-Simons (CS) theory,  with U(1) gauge symmetry, coupled to an external electromagnetic
field $\tilde{A}$ controlled by  the following effective action
\begin{equation}
S_{CS}=-\frac{k}{4\pi }\int_{\mathbb{R}^{1,2}}A\wedge dA+\frac{q}{2\pi }%
{\tilde{A}}\wedge dA.  \label{1}
\end{equation}%
Here  $A_{\mu }$ is the dynamical gauge field and $q$ is the
electron charge.  Following the Susskind approach and searching  for
extended models, it is not difficult to see that the most general
fractional quantum Hall systems,  including (\ref{1}),  can be given
by an effective gauge given by the  action
\begin{equation}
\begin{tabular}{ll}
$S= \frac{1}{4\pi }\int_{\mathbb{R}^{1,2}}[ \sum_{i,j}K_{ij}A^{i}\wedge dA^{j}+2\sum_{i}q_{i}%
\tilde{A}\wedge dA^{i}],$ &
\end{tabular}
\label{hd}
\end{equation}%
where now $K_{ij}$\ is a real, symmetric and invertible matrix
($\det K\neq 0
$).  $q_{i}$ is a vector of charges. The emergence of the $K_{ij%
}$\ matrix and the $q_{i}$ charge vector in this effective field
action are very suggestive in the sense that, besides their Lie
algebra interpretation, they can be used to embed the corresponding
 gauge theory in type II superstrings and  M-theory
compactifications.  In fact, integrating over the gauge fields
$A^{i}=dx^{\mu }A_{\mu }^{i}$ in the same way as in the  Susskind
model, one can get the   filling factor  expression
\begin{equation}
\nu =q_{i}K_{ij}^{-1}q_{j}. \label{factor}
\end{equation}
This equation  may be  thought  of as a  compact  unified
description of several kinds of fractional quantum Hall  series
recovering  the the most usual  ones. In string theory for instance,
the matrix charge $K$ encodes many geometrical and topological
properties of the singular compactified manifolds. A close
inspection shows that this matrix encapsulates  information of  the
corresponding topological invariants including the Euler
characteristic of cycles on which D-branes can wrap. The deformation
of such singularities consists on replacing the singularity by a
collection of intersecting complex curves, identified with one
complex dimensional projective  ${\bf CP}^1$. In this way, the
intersection matrices of such complex curves, up to some details,
are given  the ADE Cartan matrices. This link has been based on a
nice correspondence between the ADE root systems and ${\bf CP}^1$'s
used in the deformation of ADE singularities. Concretely, to each
simple root $\alpha_i$, one associates  a single complex curve ${\bf
CP}^1_i$. This  connection between  ADE singularities and Lie
algebras has been used to embed FQHE in string theory and related
models using D-brane physics\cite{5,6,7,8,9}.

Toric geometry, used in string theory, reveals that   the   CS
matrix charges $K$ can be identified with the Mori vectors encoding
the geometric properties associated with intersection theory of the
compactified geometry. Thus, appropriate choices of $K_{ij}$ could
determine  the filling factor values obtained recently in
theoretical and experimental works.

More recently, a topological method based on knot theory has been
used to determine filling factors in FQHE senario \cite{90}. This
may open a new window  for topological calculations including the
Euler characteristic,   which has been used in various places in
physics. In particular,  it has been used in the  context of  four
dimensional black  hole by imposing constraints on the corresponding
horizon geometries\cite{900}.

The aim of this work is to  combine toric geometry and topological
invariants of M-theory compactification to   determine filling
factors of  a class of stringy  fractional quantum Hall effect
(SFQHE).
 Concretely, we
investigate CS type models from M-theory compactified on four
complex dimensional hyper-K\"{a}hler manifolds  $X^4$. The manifolds
are viewed as target spaces of  $N=4$ sigma model in two dimensions.
They  are defined as the cotangent bundle over intersecting
2-dimensional complex toric varieties $V^2$.  The corresponding
intersection geometry is  controlled by toric Cartan matrices, which
can be considered as  CS abelian gauge charges. Using a string
duality, the presented  FQHE can be derived from wrapped D6-branes
on 4-cycles interacting with R-R gauge fields. Inspired by toric
geometry,  the associated  filling factors will be given in terms of
the intersection  numbers of intersecting $V^2_i$ using Euler
characteristic calculation based on   graph theory method. These
numbers are entries of toric Cartan matrices extending the usual
ones appearing in certain Lie algebras associated with ${\bf
CP}^1$'s considered as the building block of higher dimensional
toric manifolds. Explicit bilayer models, producing some recent
 observation of  the   filling factor values \cite{70,71},  are worked out.

The organization of this paper is as follows.  In section 2,  we
briefly review  the stringy realization of FQHE systems. Combining
toric geometry and topological calculation,  section 3 proposes a
class of toric Cartan matrices as gauge charges of CS models. In
section 4, we present a M-theory description of FQHE systems using
string duality. The computation of the corresponding filling factors
is given in terms of  toric  Cartan matrices extending  the
traditional ones arising in  Lie algebras. The last section is
devoted to conclusions and some open  questions.

\section{   String theory description of FQHE }
As mentioned  in the introduction,   the  first proposed series of
the fractional quantum states was   given by Laughlin. At low
energy, this physical system  can be modeled  by a 3-dimensional
U(1) CS theory coupled to an external electromagnetic field
$\tilde{A}$.  It has been  remarked  that this system can be
engineered  using solitonic D-branes of type II superstrings
interacting with a NS-NS \mbox{B-field} \cite{1,2}. When the B-field
is turned on, a non commutative geometry description can be explored
to engineer FQHE systems \cite{200}.

 The field
action given in  (\ref{hd}) can be derived  from the
compactification of type IIA superstring theory on local K3 surfaces
identified with the ALE spaces.  More precisely, a
 type IIA brane realization of
FQHS  in terms  of D4-branes  has  been  proposed  in
\cite{5,6,7,8,9}. This has been inspired  by   the study of quiver
gauge theories arising on the world-volume of D4-branes  wrapping
intersecting 2-spheres ${\bf CP}_{i}^{1}$  according to   Dynkin
Diagrams of  certain Lie algebras.  It is recalled that on  the
world-volume of such solitonic branes lives   a 3-dimensional theory
with several U$(1)$ gauge group factors.  To get a FQHE system,   an
external gauge field should be added. For this reason, an extra
D4-brane must be incorporated  and  wraps a generic 2-cycle given by
a linear combination of ${\bf CP}_{i}^{1} $ which reads as
\begin{equation}
\left[ C_{2}\right]=\sum_{i}q_{i}\left[{\bf CP}_{i}^{1}\right],
\end{equation}%
where ${\bf CP}_{i}^{1}$ form  a basis of $H_{2}(K3,\mathbb{Z})$. It
is observed that  there are two different ways in which the D4-brane
is wrapped on each ${\bf CP}^{1}$.  In fact, this can be supported
by the fourth homotopy group $ \Pi _{4}({\bf
CP}^{1})=\mathbb{Z}_{2}$. Indeed, if we wrap  a D4-brane over such a
geometry,   one gets two possible D2-brane states. Each one is
associated with  an equivalent class of the $\mathbb{Z}_{2}$
symmetry.  It has been shown  that the wrapped D4-brane over a ${\bf
CP}^{1}$ is sensitive to the $\mathbb{Z}_{2}$ symmetry. It carries
two charges given by  $\pm 1$. A general value of the charge $\pm q$
can also  be obtained by wrapping a D4-brane $q$ times over a ${\bf
CP}_{i}^{1}$ in two possible orientations. This procedure produces
membranes playing the role of the magnetic source found in six
dimensions. These objects should be placed perpendicularly to the
uncompactified part of the world-volume of D4-branes. Usually, on
the corresponding world-volume lives $\mbox{U}(1)^{n}$ gauge theory.
In this realization, the $q_{i}$ charges play the same role as the
D6-brane ones in ten dimensions and can be interpreted as the vector
charges of D-particles living on the corresponding space-time.   It
has been realized  that the full brane system can be described by
the $\mbox{U}(1)^{n}$ CS gauge theory with the action given by
(\ref{hd}). In this way, the matrix $K_{ij}$ encodes many
 geometric data of string theory compactified on local
backgrounds through the intersection numbers of the 2-cycles
embedded in ALE spaces. Up to some details, these intersection
numbers can be identified with   the entries of  Cartan matrices of
certain  Lie algebras. According to \cite{12,13,14}, the possible
forms of the matrix $ K$ may be grouped basically into three
categories. They are classified as follows:
\begin{enumerate}
  \item finite dimensional Lie algebras,
  \item affine Lie algebras,
  \item indefinite Lie algebras.
\end{enumerate}
 This classification has been explored to engineer  the associated   classes
of FQHE systems  as reported  in \cite{6,7}.

The rest of this paper aims  to go beyond of such classes by
relaxing certain constraints on  Cartan matrices of such Lie
algebras. Concretely,   we  use toric geometry to determine filling
factors of a class of FQHE by considering  new CS  matrix charges
derived from M-theory compactification. We will refer to them as
{\it toric Cartan matrices}. The computation will be  made in the
context of M-theory compactification on  four complex  dimensional
hyper-K\"{a}hler manifolds. These manifolds  will be built  using
$N=4$ sigma model in two dimensions.

\section{Toric  Cartan matrices as CS gauge charges}
The aim of this section is to   propose  toric Cartan matrices as CS
gauge charges by combining   toric geometry and topological
invariants of the M-theory compactification. It is recalled that
toric geometry is considered as a powerful tool to build complex
Calabi-Yau manifolds explored in string theory and F-theory  in
lower dimensions \cite{14,2000,20000,2001}. In the recent years,
several toric Calabi-Yau  manifolds have been constructed producing
interesting geometries used in the elaboration of physical systems
closed to standard model physics.  Roughly, $d$-complex dimensional
toric manifold, which we denote as
$M_{\triangle }^{d},$ is obtained  by considering the $(d+r)$%
-dimensional complex spaces $C^{d+r},$ parameterized by homogenous
coordinates $\{z=(z_{1},z_{2},z_{3},...,z_{n+r})\},$ and $r$ toric
transformations  $T_{a}$  acting on the $z_{i}$'s as follows
\begin{equation}
T_{a}:z_{i}\rightarrow z_{i}\left( \lambda _{a}^{q_{i}^{a}}\right).
\end{equation}
Here,   $\lambda _{a}$'s are $r$ non zero   complex parameters. In
this method, $q_{i}^{a}$  are integers called Mori vectors encoding
several geometrical  and topological  information on the
corresponding  manifold. These data have been used and explored in
many physical problems including in the ones appearing  in string
theory and related topics. The powerful point of this method  is the
graphic representation. In this way, the manifold is generally
represented by an integral polytope $\Delta $, known by toric
diagram,   spanned by $(d+r)$ vertices ${v}_{i}$ of the standard
lattice $Z^{d}$. In this way, the relevant pieces are the toric data
$\{v_i,q^a_i\}$. In fact, they should satisfy the following $ r$
relations
\begin{equation}
\sum_{i=1}^{d+r}q_{i}^{a}{v}_{i}=0,\qquad a=1,\ldots,r.
\end{equation}
A close inspection shows that the Mori vectors $q_i^a$    find many
places not only in  in mathematics, but also in  physics. We quote
some possible links:
\begin{enumerate}
  \item in connection with two  dimensional
field theory,   the $q_{i}^{a}$ integers are interpreted, in the
${\cal N}=2$ gauged linear sigma model as the $\mbox{U(1)}^{r}$
gauge charges of ${\cal N}=2$ chiral multiplets,
  \item  from intersection  theory,   they have also a
nice geometric  interpretation in terms of the intersections of
complex curves $C_{a}$ and divisors $D_i$ of $M_{\triangle}^{d}$
\cite{14},
  \item  they  encode  also  information on topological invariants
  including the Euler characteristic of the corresponding  middle cohomology.
\end{enumerate}
In lower dimensional toric manifolds, including the  K3 surfaces,
the two last links  are related to  Lie algebras. Deleting the non
compact divisors, the Mori vectors, up some details, can be
identified with Cartan matrices of certain Lie algebras.   To see
the relation with toric geometry,  it is useful  to recall that for
a semi-simple Lie algebra,  the Cartan matrices usually denoted by
$K$ have  the following properties:
\begin{enumerate}
  \item the diagonal entries $K_{ii}$
take  the same positive integer 2,
    \item all off-diagonal entries $K_{ij}$  can only  take  the non-positive
    integers,
  \item the possible  non-positive integers  are $-3, -2, -1, 0$
  \item $K_{ij}=0$ implies  $K_{ji}=0$.
\end{enumerate}
In fact, one should distinguish three types of  Lie algebras
classified in terms of $K$ as follows
\begin{equation}
\begin{tabular}{llll}
finite Lie algebras & : & $ dek\; K>0$ &  \\
affine Lie algebras & : & $ det\;K=0$ &  \\
indefinite Lie algebras & : & $det\; K<0.$ &
\end{tabular}%
\end{equation}
Following \cite{13,14}, finite  Lie algebras are used in the
engineering of non abelian gauge theories while affine   are
infinite dimensional ones playing   an important role in  dealing
with  conformal invariance and critical phenomena. The indefinite
sector is the biggest and its role is still unclear in physics.
There are few works dealing with   the corresponding  field
theoretical realization.

In string theory, the most important application  of  Cartan
matrices is the relation with  the  intersection of the complex
curves inside the K3 surfaces on which D-branes can wrap to produce
quiver gauge theories.  Deleting non compact divisors, the
intersection matrix for two complex curves ${\bf CP}^1_i$ and ${\bf
CP}^1_j$ can be identified with Cartan matrices of the above Lie
algebras. These intersections $[{\bf CP}_i^1].[{\bf CP}^1_j]=K_{ij}$
are given by
\begin{equation}
K_{ij}=2\delta_{ij}-I_{ij}.
\end{equation} The
diagonal entries describing  the self intersection of one
dimensional projective space ${\bf CP}^1_i$ reads as
\begin{equation}
K_{ii}=\chi ({\bf CP}_i^1)=2,
\end{equation}
where  $\chi ({\bf CP}_i^1)$ denotes  their Euler characteristics.
$I_{ij}$ depends on the geometry  in  question. It is recalled that
in the case of $A_n$ singularity, this matrix takes the following
form
\begin{equation}
I_{ij}=\delta_{ij-1}+\delta_{ij+1}.
\end{equation}
Motivated from the fact that  ${\bf CP}^1$  is the building block of
higher dimensional of toric manifolds,  we will present   models
going beyond  such a link, which  will be explored in the
calculation of the filling factors of a class of FQHE. In
particular, we use toric geometry to compute such  factors  in the
context of  M-theory compactification.  In fact, M-theory motives us
to consider manifolds $X$ with  the following relaxing condition
\begin{equation}
\chi (X)= 2,
\end{equation}
In connection with topological calculation,  it is recalled that the
Euler characteristic  can be obtained from graph theory associated
with the geometric  triangulation.  For a planar graph theory, this
quantity takes the form
\begin{equation}
\chi (X)=v-e+f
\end{equation}
where $v$, $e$, $f$  indicate   the number of vertices, edges and
faces respectively \cite{2002}. In what  follows, we  consider
geometries with the following constraints
\begin{equation}
v-e+f\neq 2.
\end{equation}
A priori there  are many    geometries satisfying   such a
topological condition. However, M-theory compactification requires
us to consider higher dimensional toric manifolds   as  a possible
generalization of ${\bf CP}^1$.  The latter  is
  defined by $ r = 1$ and the vector charge $q_i = (1,1)$. It is represented  by two vertices $v_1$ and $v_2$
    satisfying the following constraint toric equation
\begin{equation}
v_1+v_2=0.
\end{equation}
In  toric geometry language, ${\bf CP}^1$  is represented by a
toric graph identified with an interval $[v_1,v_2]$
  with
a circle on top.  The latter  vanishes at the end points $v_1$ and
$v_2$ \cite{14,2000,2001}.

We will see later that  in  the embedding of  FQHE in M-theory, the
complex curve ${\bf CP}^1$,   used in previous section,  should be
replaced by a two dimensional toric variety $V^2$. The latter  can
be represented by a toric diagram (polytope) $\Delta(V_2 )$ spanned
$=2+r$ vertices $v_i$ of the $Z^2$ lattice satisfying  the following
toric constraints
\begin{equation}
\sum\limits_{i=1}^{n}q_i^av_i=0, \qquad a=1,\ldots,r, \qquad n=2+r
\end{equation}
where $q^a_i$   are the corresponding Mori vectors.  This class of
manifolds includes  del Pezzo and Hirzebruch complex surfaces.
However, the simple example is ${\bf CP}^2$ defined by $r = 1$ and
the Mori vector charge $q_i = (1, 1, 1)$. The corresponding toric
diagram has 3 vertices $v_1$, $v_2$ and $v_3$. Its polytope, which
is a finite sublattice of the $Z^2$ square lattice,  is described by
the intersection of three ${\bf CP}^1$ curves. It  defines a
triangle $(v_1v_2v_3)$ in the $R^2$ plane constrained by
\begin{equation}
v_1+v_2+v_3=0.
\end{equation}

In what follows, we combine toric geometry and topological
invariants to compute  filling factors of   FQHE  embedded  in
M-theory compactification. More concretely, we consider the
following  toric Cartan matrices
\begin{equation}
K_{ij}=[V_i^2].[V^2_j].
\end{equation}
They will be identified as CS gauge charges to compute the
corresponding filling factors. As  in the case of ${\bf CP}^1$'s,
the intersection numbers should  take the following form
\begin{equation}
[V_i^2].[V^2_j]=\chi(V^2)\delta_{ij}-I_{ij},
\end{equation}
which will be argued in the coming sections.
  \section{    Filling factors  of    FQHE   from  M-theory  toric geometry }
  In this section, we investigate   a  class of quiver gauge theories describing multilayered  states in
FQHE from toric geometry used  in string theory compactification. In
this way, the  corresponding matrix charges can take values which
are different from the ones appearing in the Cartan matrices of Lie
algebras (finite, affine and indefinite) associated with $\bf CP^1$.
Instead of having $K_{ii}=\chi({\bf CP}^1)=2$, we will consider
arbitrary integers associated with the Euler characteristic of two
dimensional toric varieties $V^2$. To start, we recall first that
very little is known about the application of such matrices in
string theory compactification. For this reason, the general study
is beyond the scope of the present work, though we will consider
some examples dealing with  three dimensional CS gauge theories with
toric Cartan matrices given by (16), as matrix gauge charge. We will
see that  these models can be embedded in M-theory on four complex
dimensional hyper-K\"{a}hler manifolds $X^4$ associated with the
following decomposition
\begin{equation}
{\mathbb{R}^{1,10}}\to {\mathbb{R}^{1,2}}\times X^4.
\end{equation}
Here,   $ X^4$ will be identified with the cotangent bundle over
intersecting two dimensional toric varieties $V^2$. A  very nice way
to discuss such geometries is to use the technics of $N=4$ sigma
model in two dimensions and its relation to toric geometry. In this
way, the geometries, that will be used,  are realized as the moduli
space of $N=4$ supersymmetric $\mbox{U}(1)^r$  gauge theory in two
dimensions with $(r+2)$ hyper-multiplets. These physical ingredients
give the dimension of the space $(4(r+2-r) = 8)$  on which M-theory
will be compactified\cite{21}. The corresponding geometries are
solutions of the following D-flatness condition
\begin{equation}\label{sigma4}
\sum_{i=1}^{r+2}Q_{i}^{a}[\phi _{i}^{\alpha }{\bar{\phi}}_{i\beta
}+\phi _{i}^{\beta }{\bar{\phi}}_{i\alpha
}]=\vec{\xi}_{a}\vec{\sigma}_{\beta }^{\alpha },\;\;a=1,\ldots ,r
\end{equation}%
where $Q_{i}^{a}$ is the sigma model  matrix charge. $\phi
_{i}^{\alpha }$'s ($\alpha =1,2)$ denote the component field
doublets of each hypermultiplets ($i=1\ldots ,r+2) $.
$\vec{\xi}_{a}$ are the Fayet-Iliopolos (FI) 3-vector couplings
rotated by $\mbox{SU}(2)$  symmetry, and $\vec{\sigma}_{\beta
}^{\alpha }$ are the traceless $2\times 2$ Pauli matrices.  From the
obvious similarity with the $N=2$ sigma model describing K\"{a}hler
manifolds,  it is not surprising to have similar results for
intersecting cycles  in  four complex dimensional hyper-K\"{a}hler
manifolds. In fact, the solutions of eqs(\ref{sigma4}) depend on the
number of the gauge fields, charges and  the values of the FI
couplings. To see  how this works in practice, we  consider first
the case of $\mbox{U}(1)$ gauge theory. It is not difficult to
realize  that eqs(\ref{sigma4}) describe the cotangent bundle over a
toric variety  $V^2$ ($T^*(V^2)$). Moreover, the intersection theory
assigns the intersection number to complex surfaces inside of these
eight real  dimensional hyper-K\"{a}hler manifolds.  Regarding $V^2$
as the zero section of $T^*(V^2)$  and using toric geometry ideas,
the self-intersection number, being equal to the Euler
characteristic of $V^2$,  reads as
\begin{equation}
[V^2].[V^2]=\chi(V^2)=v-e+f,
\end{equation}
Combining toric geometry and topology, we can compute  such a
quantity.  For simplicity reason, we consider  a class of  $V^2$
having a planar  toric  graph. Identifying such a toric graph with
the one used for computing the Euler characteristic, we can compute
$\chi(V^2)$. For some physical reason, we will consider toric
manifolds   with the following constraint
\begin{equation}
e=f.
\end{equation}
This  constraint can be understood  as a duality  in two dimensions.
In this way, the above equation reduces to
\begin{equation}
\chi(V^2)=n,
\end{equation}
 where $n$ is  no longer restricted to 2 as  in the case of one dimensional toric  ${\bf CP}^1$
associated with Cartan  matrices of certain  Lie algebras. It
describes the number of vertices of  $\Delta(V_2 )$.  In the general
case corresponding to the $\mbox{U}(1)^r$ gauge theory,
eqs(\ref{sigma4}) describe the cotangent bundle over $r$
intersecting $V^2$'s.  This geometry may extend the intersecting
2-cycles  in  ALE spaces classified by  Lie algebras used  in the
geometric engineering method of the quantum field theory
\cite{13,14}. Assuming that $V^2_{i}$ intersects $V^2_{i+1}$ at
$CP^1$,  the intersection numbers of $V_i^2$ read as
\begin{equation}
\lbrack V^2_{i}]\cdot \lbrack V^2_{i}]=n,\qquad \lbrack
V^2_{i}]\cdot \lbrack V^2_{i+1}]=-2.
\end{equation}%
This shows that the  self-intersection number of   $V^2_{i}$ is $n$
and $V^2_{i}.V^2_{j}=0$
   $V^2_{i}$  does not intersect $V^2_{j}$.

Having constructed the M-theory background, we  move now to get the
corresponding  CS gauge theories. Using M-theory/Type IIA duality,
the $\mbox{U(1)}^r$ CS models controlled by the action (2)  can be
derived  from D6-branes wrapping on such intersecting geometries
filling the 3-dimensional Minkowski space on which  FQHE will
reside. To get  an effective theory of a  hierarchical description,
let us first start with a CS theory with a single U(1) gauge
symmetry. In this realization,  the WZ action, associated with a
D6-brane, reads as
\begin{equation}
S_{WZ}\sim \int_{\mathbb{R}^{1,6}} F\wedge F\wedge A_3+\int_{\mathbb{R}^{1,6}} F\wedge F\wedge F\wedge A_1
\end{equation}
 where  $A_1$ and $A_3$ are  the RR 3-form
coupled to the D0-brane and D2-brane respectively. Integrating by
part and ignoring the  D0-brane contributions, the D6 brane WZ
action can be reduced to
\begin{equation}
\int_{\mathbb{R}^{1,6}} F\wedge F\wedge A_3=-\int_{\mathbb{R}%
^{1,6}} A\wedge F\wedge (dA)_4.
\end{equation}
Integrating over one single  ($V^2$), one can obtain   the first CS
term of eq.(\ref{1}) namely
\begin{equation}
-\int_{\mathbb{R}%
^{1,2}\times V^2} A\wedge F\wedge (dA)_4\to -\frac{k}{4\pi}
\int_{\mathbb{R}^{1,2}} A\wedge F,
\end{equation}
where   $k$ is  now identified with
$\frac{1}{2\pi}\int_{V^2}(dA)_4$. To couple the system to an
external magnetic  field, we  should  add the RR 5-form $A_5$
sourced by a solitonic  D4-brane being dual to a D2-brane.  The
decomposition  of this field gives
\begin{equation}
A_5\to \tilde A \wedge \omega.
\end{equation}
It is worth noting that  $\tilde A$ is  an U(1) gauge field which
can be obtained from the dimensional reduction of the RR 5-form on
$V^2$ while  $\omega$ is a harmonic 4-form on  it. In this
representation, the WZ term $\int A_5\wedge F$ on a D6-brane
produces the following term
\begin{equation}
q \int_{\mathbb{R}^{1,2}} \tilde A \wedge F.
\end{equation}
The   gauge field   $\tilde A$   can be interpreted now  as a
magnetic external gauge field which  couples to the studied  FQHE
systems. This analysis can be extended  to   a
 CS gauge theory with several  $\mbox{U}(1)$  gauge factors.  This  system
can be modeled  from a stack of D6-branes wrapping individually
intersecting  the  $V^2$'s.  Such 4-cycles    can be  realized as
the middle cohomology of $X^4$.  In this way, the external magnetic
source   can be identified  by a linear combination of $V_{i}^{2} $
which reads as
\begin{equation}
\left[ C_{4}\right]=\sum_{i}q_{i}\left[V_{i}^{2}\right],
\end{equation}%
where $V_{i}^{2}$ denote a basis of $H_{4}(X^4,\mathbb{Z})$.

In  the rest, we present concrete examples  corresponding  to the
famous values of the filling factor of    bilayer systems associated
with $\mbox{U(1)} \times \mbox{U(1)}$ quiver CS gauge theory.  We
treat    a particular  value associated with $\nu=\frac{2}{5}$. In
fact, the value $\nu=\frac{6}{5}= \frac{3\times 2}{5}$ can be
obtained by considering  a bilayer system  with $\mbox{U(1)}\times
\mbox{U(1)}$
 quiver gauge theory  from   two D6-branes wrapping on two intersecting projective
spaces ${\bf CP}^2$'s with   the following   toric  Cartan  matrix
 \begin{equation*}
K_{ij}(\mbox{U}(1)\times \mbox{U}(1))=\begin{pmatrix}
3 & -2 \\
-2 & 3
\end{pmatrix}.
\end{equation*}
The vector charge $q_i=(1,0)$  recovers the above value of the
filling factor.  The second example concerns a bilayer system
corresponding to  $\mbox{U(1)}\times \mbox{U(1)}$  quiver gauge
theory with the following toric  Cartan  matrix
 \begin{equation*}
K_{ij}(\mbox{U}(1)\times \mbox{U}(1))=\begin{pmatrix}
n & -2 \\
-2 & n
\end{pmatrix}.
\end{equation*}
This matrix is associated with intersecting  two  generic toric
manifold $V^2$, with $n$ vertices,  at $CP^1$. It is recalled that
this includes the so called  del Pezzo surface obtained by  blowing
up  $\bf CP^2$. To avoid toric geometry problems, we consider only
toric ones.  Evaluating (3) for the charges $q_i = (1, 0)$, we get
\begin{equation}
\nu=\frac{n}{n^2-4}.
\end{equation}
This is a general expression containing some known  values.  In
particular, taking $n= 3$, one recovers again  the famous value
$\frac{3}{5}$.
\section{ More on the filling factor expressions}
Other interesting  filling factor values  can be  modeled by
considering complicated geometries associated
 with connected  sum of complex curves. Taking  a connected sum of $n$ torii $T^2$, denoted
 $\Sigma_n$,
  the corresponding Euler
 characteristic  reads as
\begin{equation}
\chi(\Sigma_n)=2(1-n)
\end{equation}
Assuming  that the intersection, in M-theory compactification,   is
homotopic to such a connected sum and after a  scaling factor,  we
expect to have the following intersecting number relations
\begin{equation}
\lbrack V^2_{i}]\cdot \lbrack V^2_{i}]=n,\qquad \lbrack
V^2_{i}]\cdot \lbrack V^2_{i+1}]=1-n.
\end{equation}
 It is noted  that the  subsequence of the Jain's series can
be derived  using  such intersection numbers. Indeed,  the filling
factor value
\begin{equation*}
\nu=\frac{n}{2n-1},
\end{equation*}
can be obtained  by taking  the components of the vector charge  as
1 and 0 and the following  toric Cartan  matrix charge
\begin{equation*}
K_{ij}(\mbox{U}(1)\times \mbox{U}(1))=\begin{pmatrix}
n & 1-n \\
1-n & n
\end{pmatrix}.
\end{equation*}
The  corresponding  filling factor  can be derived from two
D6-branes wrapping on two intersecting  toric varieties represented
by a toric diagram having $n$ vertices.   Using the fact that
\begin{equation*}
\nu=\nu_1+\nu_2=\frac{n}{2n-1},
\end{equation*}
where
\begin{equation*}
\nu_1=\frac{n-2}{2n-1},  \qquad\nu_2=\frac{2}{2n-1},
\end{equation*}
associated with   single layer FQH states, we can make contact with
recent observed values. Indeed, the first filling factor
$\nu_1=\frac{n-2}{2n-1}$ contains some experimentally observed
fractional values of the filling factor \cite{70,71}. The second
filling factor $\nu_2=\frac{2}{2n-1}$ involves also some known
series used in the study of the  trial functions explored in the
study of FQHE appearing in the graphene material physics. In fact,
these types of solutions have been proposed first by Laughlin in
\cite{27}, which
 have been generalized by Halperin. This has  been used  in order to
 deal with multi-components with SU(K) local symmetries \cite{28}. Inspired by
the Halperin analysis on the study of the  trial wave functions
$(m_1, m_2, m_3)$ of QHE, we can recover some result corresponding
to  the graphene physics. Concretely,  the value
$\nu_2=\frac{2}{2n-1}$  has been found   in the study of the $(n, n,
n-1)$ trial wave functions given in \cite{29}. It is interesting to
note that one can also recover Laughlin's wave functions.

\section{Conclusion and discussion}

In this work, we  have built  a  SFQHE  with toric Cartan matrices
as abelian CS  gauge charges. More precisely, we have investigated
CS type models from M-theory compactified on four complex
dimensional hyper-K\"{a}hler manifolds $X^4$. The corresponding
manifolds are viewed as target spaces of $N=4$ sigma model in two
dimensions. In particular,  we have considered the cotangent bundle
over intersecting 2-dimensional complex toric varieties $V^2$
associated with toric  Cartan matrices. Using  string/M-theory
duality, FQHS can  be obtained from wrapped D6-branes  interacting
with R-R gauge fields. We have interpreted the corresponding filling
factors in terms of the intersections between   two dimensional
toric varieties $V^2_i$. Toric geometry and Euler characteristic
topological calculations show that the corresponding intersection
numbers are given in terms of toric Cartan matrices which  can be
thought of  as a generalization class of Cartan matrices of Lie
algebras, associated with the intersection of $\bf CP^1$'s. Explicit
models producing some known filling factor values have been
elaborated. In particular, we have analyzed in some detail the cases
of bilayer systems. Using toric geometry, we have shown how the
physical constraints on the corresponding M-theory on backgrounds
can be related to Jain series.

This bilayer calculation could be  pushed further by considering
multilayered systems. In this way, the filling factor expressions
depend on two relevant parameters $n$ and $r$, namely
\begin{equation*}
\nu=\nu(n,r)
\end{equation*}
where $r$ is  the  CS matrix charge size. For  $r=3$ associated with
a  tri-layer system, the matrix charge can take the following form
 \begin{equation*}
K_{ij}=\begin{pmatrix}
n & -2 & 0\\
-2 & n & -2 \\
0 & -2 & n
\end{pmatrix}.
\end{equation*}
 Evaluating (3) for the charges $q_i = (1, 0,0)$, we  obtain
\begin{equation*}
\nu(n,3)=\frac{n^2-4}{n^3-8n},
\end{equation*}
which could produce new  filling factor values.

 This work comes up
with many open questions. A natural one concerns the connection with
Borcherds and Monster Lie algebras. The latters  are considered as
generalized Kac-Moody algebras.  We hope to report on this issue in
future works.


\begin{thebibliography}{99}

\bibitem{1} B. A. Bernevig, J. Brodie, L. Susskind and N. Toumbas, {\em How Bob Laughlin Tamed the Giant
Graviton from Taub-NUT space}, JHEP{\bf 0102}(2001)003, {\tt
hep-th/0010105}.

\bibitem{2}
O. Bergman, {\em  Quantum Hall Physics in String Theory}, {\tt
hep-th/0401106}.
\bibitem{200}
 L. Susskind,  {\em The Quantum Hall Fluid and
Non-Commutative Chern Simons Theory},  {\tt arXiv:hep-th/0101029}.


\bibitem{3}
M. Fujita, W. Li, S. Ryu, T. Takayanagi, {\em Fractional Quantum
Hall Effect via Holography: Chern-Simons, Edge States, and
Hierarchy}, JHEP {\bf 0906} (2009)066, {\tt
arXiv:0901.0924[hep-th]}.
\bibitem{4} B. Roy, M.  P. Kennett, S. D.  Sarma,  {\em
Chiral Symmetry Breaking and the Quantum Hall Effect in Monolayer
Graphene},  Phy.  Rev.  {\bf B90} (2014)201409(R) , {\tt
arXiv:1406.5184}
\bibitem{5}
A. Belhaj, A. Segui, {\em  Engineering of Quantum Hall Effect from
Type IIA String Theory on The K3 Surface}, Phys. Lett. {\bf B691}
(2010)261-267,  {\tt arXiv:1002.2067[hep-th]}.

\bibitem{6} A. Belhaj et al. {\em  Embedding Fractional Quantum Hall Solitons
in M-theory Compactifications},  Int.J.Geom.Meth.Mod.Phys. {\bf 8}
(2011) 1507-1518   {\tt arXiv:1007.4485[hep-th]}.
\bibitem{7} A. Belhaj et al. {\em  Brane Realizations of Quantum Hall Solitons and
 Lie Algebras},  Int.J.Geom.Meth.Mod.Phys. {\bf 9} (2012) 1250017.







 \bibitem{8}A. Belhaj, {\em On Fractional Quantum
Hall Solitons and Chern-Simons Quiver Gauge Theories},
Clas.Quant.Grav. {\bf 29} (2012) 095013 , {\em arXiv:1111.1878}.

 \bibitem{9}
A. Belhaj,  {\em  On Fractional Quantum Hall Solitons in ABJM-like
Theory}, Phys.Lett. {\bf B705} (2011) 539-542  {\tt
arXiv:1107.2295}.

 \bibitem{90}
 C. Pe\~{n}a, {\em Determining the filling factors of fractional quantum Hall
states using knot theory}, {\tt  arXiv:1504.03645}.


\bibitem{10}
R. B. Laughlin, {\em  Anomalous Quantum Hall Effect: An
Incompressible Quantum Fluid with Fractionally Charged Excitations},
Phys. Rev. Lett.{\bf 50}(1983)1395.
\bibitem{11}
X-G. Wen, Quantum Field Theory of Many-body Systems, Oxford
University Press, 2004.
\bibitem{900}
R Aros, D E Diaz, A Montecinos, {\em Logarithmic correction to BH
entropy as Noether charge}, JHEP {\bf 1007}(2010)012, {\tt
arXiv:1003.1083}.
\bibitem{70}
  X. Lin, R. R.  Du, X. Xie,  {\em Recent Experimental
Progress of Fractional Quantum Hall Effect: 5/2 Filling State and
Graphene}, National Science Review,1(2014)564-579, {\tt
arXiv:1501.00073}

\bibitem{71}
N. Samkharadze, I. Arnold, L.N. Pfeiffer, K.W. West, G.A. Cs\'athy,
{ \em Observation of Incompressibility at $\nu=4/11$ and
$\nu=5/13$}, {\tt arXiv:1412.8186}.
\bibitem{12}
V. G. Kac, Infinite dimensional Lie algebras, third edition,
Cambridge University Press (1990).
\bibitem{13}
A. Belhaj, M. B . Sedra, {\em   Quiver Gauge theories from Lie
Superalgebras},   Afr.Rev.Phys.{\bf  9} (2014) 311, {\tt
arXiv:1207.657}.
\bibitem{14}
S. Katz, P. Mayr, C. Vafa, {\em Mirror symmetry and exact solution
of 4d N = 2 gauge theories I}, Adv. Theor. Math. Phys. {\bf 1}
(1998) 53, {\tt hep-th/9706110}.
\bibitem{2000}
 W. Fulton, {\em Introduction to Toric varieties}, Annals of Math. Studies, {\bf No .131}, Princeton
University Press, 1993.
\bibitem{2001} N.C. Leung and C. Vafa, {\em  Branes and toric
geometry}, Adv .Theo. Math. Phys {\bf 2}(1998) 91, {\tt
hep-th/9711013}.
\bibitem{20000} Y.
Aadel, A. Belhaj, Z. Benslimane, M. B. Sedra, A. Segui, {\em  Qubits
from Adinkra Graph Theory via Colored Toric Geometry}, {\tt
arXiv:1506.0252}.
\bibitem{2002} M. Nakahara,   {\em Geometry, Topology and Physics},
Second Edition.
\bibitem{21}
A. Belhaj, {\em  Manifolds of G2 Holonomy from N=4 Sigma Model}, J.
Phys. {\bf  A35} (2002) 8903,  {\tt hep-th/0201155}.
\bibitem{27} R. B. Laughlin,  {\em Anomalous Quantum Hall Effect: An
Incompressible Quantum Fluid with Fractionally Charged Excitations},
Phys. Rev. Lett. {\bf 50}(1983)1395.
\bibitem{28} B. I. Halperin, Helv. Phys. Acta
{\bf 56} (1983)75.
\bibitem{29} Z. Papica, M. O. Goerbiga, N. Regnault, {\em  Theoretical expectations
for a fractional quantum Hall effect in graphene}, {\tt
arXiv:0902.3233}.


\end{thebibliography}
\end{document}